\documentclass[10pt]{article}
\usepackage[english]{babel}

\usepackage{amsthm}
\usepackage{amsmath}
\usepackage{hyperref}
\usepackage{graphicx} 
\usepackage[utf8]{inputenc}
\usepackage{subcaption}
\usepackage{tikz}
\usetikzlibrary{bayesnet}
\usetikzlibrary{positioning}
\usetikzlibrary{decorations.text}
\usetikzlibrary{decorations.pathmorphing}
\usetikzlibrary{backgrounds}
\usepackage{bm}
\usepackage{multirow}
\usepackage{amssymb}
\usepackage{float}
\usepackage{units}

\usepackage{multirow}
\usepackage[utf8]{inputenc} 
\usepackage[T1]{fontenc}  
\usepackage{adjustbox}
\usepackage[margin=1cm]{caption}

\usepackage[b5paper, twoside]{geometry}
\usepackage{titlesec}

\usepackage[longnamesfirst, round]{natbib}
\setcitestyle{authoryear,open={(},close={)}}

\numberwithin{equation}{section}
\theoremstyle{plain}

\theoremstyle{definition}

\definecolor{basecolor}{HTML}{C4C4C4}

\usepackage{listings}
\usepackage{xcolor}
\definecolor{codegreen}{rgb}{0,0.6,0}
\definecolor{codegray}{rgb}{0.5,0.5,0.5}
\definecolor{codepurple}{rgb}{0.58,0,0.82}
\definecolor{backcolour}{rgb}{0.95,0.95,0.92}

\lstdefinestyle{mystyle}{
    backgroundcolor=\color{backcolour},   
    commentstyle=\color{codegreen},
    keywordstyle=\color{magenta},
    numberstyle=\tiny\color{codegray},
    stringstyle=\color{codepurple},
    basicstyle=\ttfamily\footnotesize,
    breakatwhitespace=false,         
    breaklines=true,                 
    captionpos=b,                    
    keepspaces=true,                 
    numbers=left,                    
    numbersep=5pt,                  
    showspaces=false,                
    showstringspaces=false,
    showtabs=false,                  
    tabsize=2
}

\title{GeoAdjust: Adjusting for Positional Uncertainty in Geostatistial Analysis of DHS Data}

\date{}

\author{Umut Altay\qquad John Paige\qquad Andrea Riebler\qquad \\Geir-Arne Fuglstad\\ \\Department of Mathematical Sciences, Norwegian University \\of Science and Technology, Trondheim, Norway}

\begin{document}
\maketitle

\abstract{
The R-package \href{https://github.com/umut-altay/GeoAdjust-package}{\tt GeoAdjust} implements fast empirical Bayesian geostatistical inference for household survey data from the  Demographic and Health Surveys Program (DHS) using Template Model Builder (TMB). DHS household survey data is an important source of data for tracking demographic and health indicators, but positional uncertainty has been intentionally introduced in the GPS coordinates to preserve privacy. \href{https://github.com/umut-altay/GeoAdjust-package}{\tt GeoAdjust} accounts for such positional uncertainty in geostatistical models containing both spatial random effects and raster- and distance-based covariates. The R package supports Gaussian, binomial and Poisson likelihoods with identity link, logit link, and log link functions respectively. The user defines the desired model structure by setting a small number of function arguments, and can easily experiment with different hyperparameters for the priors. \href{https://github.com/umut-altay/GeoAdjust-package}{\tt GeoAdjust} is the first software package that is specifically designed to address positional uncertainty in the GPS coordinates of point referenced household survey data. The package provides inference for model parameters and can predict values at unobserved locations.
}

\section{Introduction}

In each demographic health survey collected by the DHS program, positional uncertainty is intentionally introduced in the GPS coordinates of the  household cluster centers as a privacy protection measure\citep{DHSspatial07}. 

The random displacement procedure, or \emph{jittering} scheme, is publicly known \citep{DHSspatial07}. The jittering can be an issue because traditional geostatistical analyses assume locations are known exactly, and we have recently shown that ignoring the positional error in DHS data may lead to attenuated estimates of the covariate effect sizes and reduced predictive performance \citep{altay2022accounting,altay2022covariates}. 

While common practice is to ignore jittering, some approaches have been put forward to account for it. With respect to the error induced in spatial covariates, \citet{warren2016influenceOne} proposed regression calibration for distance-based covariates, and \citet{perez2013guidelines,perez2016influence} proposed using a 5 km moving window (or buffer zone) for raster-based covariates. However, these approaches do not address the attenuation arising in the covariate effect sizes when replacing the true covariate with a proxy. With respect to the error induced in the spatial effect, \citet{fanshawe2011spatial} proposed a Bayesian approach in the limited setting of no covariates and Gaussian observation model. \citet{wilson2021estimation} proposed a more complex approach using INLA-within-MCMC \citep{rue2009approximate, gomez2018markov}, which could handle the error induced in both the spatial random effect and in spatial covariates, but computation time is too extensive for routine use of the approach. None of the mentioned papers provide an R package for easy application of the methods.

With the package \href{https://github.com/umut-altay/GeoAdjust-package}{\tt GeoAdjust} we address the need for fast, flexible and user-friendly software to estimate geostatistical models for DHS data subject to positional uncertainty. \href{https://github.com/umut-altay/GeoAdjust-package}{\tt GeoAdjust} addresses the positional uncertainty by adjusting for jittering both in the spatial random effect and spatial covariates, and achieves fast inference by combining the computational feasibility of the stochastic partial differential equations (SPDE) approach \citep{Lindgren:etal:11} with the autodifferentiation feature of the Template Model Builder (TMB) R-package \citep{JSSv070i05}. The R-package GeoAdjust is on CRAN \citep{Rmain} and can be installed by install.packages("GeoAdjust") command.

\section{Geostatistical inference under jittering}
We consider a country with spatial domain $\mathcal{D}\subset\mathbb{R}^2$, where $C$ small groups of households, called \emph{clusters}, are observed. For clusters $c = 1, \ldots, C$, we denote the true location by $\boldsymbol{s}_c^* \in\mathcal{D}$, and we denote the observed location, provided by DHS, by $\boldsymbol{s}_c \in\mathcal{D}$. Additionally, each cluster has a known classification as urban (U) or rural (R). The observed locations are linked to the true locations via a known jittering distribution $\pi_{\mathrm{Urb}[c]}(\boldsymbol{s}_c|\boldsymbol{s}_c^*)$. The subscript $\mathrm{Urb}[c]\in\{\mathrm{U},\mathrm{R}\}$ is necessary since the DHS uses different jittering distributions in urban and rural clusters. Urban clusters are jittered up to $2\, \mathrm{km}$, and rural clusters are jittered up to $5\, \mathrm{km}$ with probability $0.99$ and jittered up to $10\, \mathrm{km}$ with probability $0.01$ \citep{DHSspatial07}. The angle and jittering distance are sampled from uniform distributions, but the boundaries of either the first or the second administrative level are respected.

We model responses $y_1, \ldots, y_C$ and observed locations $\boldsymbol{s}_1, \ldots, \boldsymbol{s}_C$ jointly as
\begin{align}
y_c \mid \eta_c, \boldsymbol{\phi} &\sim \pi(y_c \mid \eta_c, \boldsymbol{\phi}), \quad \boldsymbol{s}_c|\boldsymbol{s}_c^*\sim \pi_{\mathrm{Urb}[c]}(\boldsymbol{s}_c|\boldsymbol{s}_c^*), \nonumber\\
      \eta_c &= \eta(\boldsymbol{s}_c^*),\label{eq:model}
\end{align}
for $c = 1, \ldots, C$, where  $\pi(y_c \mid \eta_c, \boldsymbol{\phi})$ is the likelihood of $y_c$ with linear predictor $\eta_c$ and likelihood parameter vector $\boldsymbol{\phi}$, and $\eta(\cdot)$ is a Gaussian random field describing spatial variation. The linear predictor is linked to the mean of the likelihood family through a link function. The package implements the identity link in the case of a Gaussian likelihood, the log-link for Poisson and the logit-link for the binomial likelihood.

The latent spatial variation is modelled as
\[
    \eta(\boldsymbol{s}^*) = \boldsymbol{x}({\boldsymbol{s}_c^*)}^\mathrm{T}\boldsymbol{\beta}+u({\boldsymbol{s}_c^*}), \quad \boldsymbol{s}^* \in\mathcal{D},
\]
which combines $p$ spatial covariates, $\boldsymbol{x}(\cdot)^\mathrm{T}$, with a Matérn Gaussian random field (GRF), $u(\cdot)$, with fixed smoothness $\nu = 1$.
The coefficients of the covariates are assigned a Gaussian prior $\boldsymbol{\beta}\sim\mathcal{N}_p(\boldsymbol{0},V \mathbf{I}_p)$, where $V$ is a fixed variance. The parameters of the Matérn GRF, the spatial range $\rho_\mathrm{S}$ and marginal variance $\sigma_\mathrm{S}^2$, are assigned penalized complexity (PC) priors with $\mathrm{P}(\rho_\mathrm{S} > \rho_0) = 0.50$ and $\mathrm{P}(\sigma_\mathrm{S} > 1) = 0.05$ \citep{fuglstad:etal:19a}. We recommend choosing the median range $\rho_0$ as the 10\% of the diameter of $\mathcal{D}$ to be able to capture the spatial variability at moderate distances. 

To complete the specification of the model, we need to assign a prior for the true cluster locations. We choose a uniform prior $s_c^*\sim \mathcal{U}(\mathcal{D})$ so that all $\boldsymbol{s}_c^*$ compatible with $\boldsymbol{s}_c$ are equally likely \emph{a priori}, $c = 1, \ldots, C$. More complicated choices taking population density or urban/rural status into account are possible, but such rasters would have to be estimated and could be biased and uncertain.
\href{https://github.com/umut-altay/GeoAdjust-package}{\tt GeoAdjust} treats the unknown true locations as nuisance parameters and integrates them out,
\begin{align}
    \pi(y_c, \boldsymbol{s}_c|\eta(\cdot)) &= \int_{\mathcal{D}} \pi(y_c, \boldsymbol{s}_c| \eta(\cdot), \boldsymbol{s}_c^*) \pi(\boldsymbol{s}_c^*) \ \mathrm{d}\boldsymbol{s}_c^* \notag \\
    &= \int_{\mathcal{D}} \pi(y_c| \eta(\boldsymbol{s}_c^*)) \pi_{\mathrm{Urb}[c]}(\boldsymbol{s}_c| \boldsymbol{s}_c^*) \pi(\boldsymbol{s}_c^*) \ \mathrm{d}\boldsymbol{s}_c^*.\label{eq:numInt}
\end{align}
 
We use the SPDE approach \citep{Lindgren:etal:11} to describe $u(\cdot)$ and use the speed and flexibility of autodifferentiation in TMB to perform inference quickly.

\section{Package structure and functionality}
\begin{figure}
\centering
\includegraphics[width=.85\textwidth]{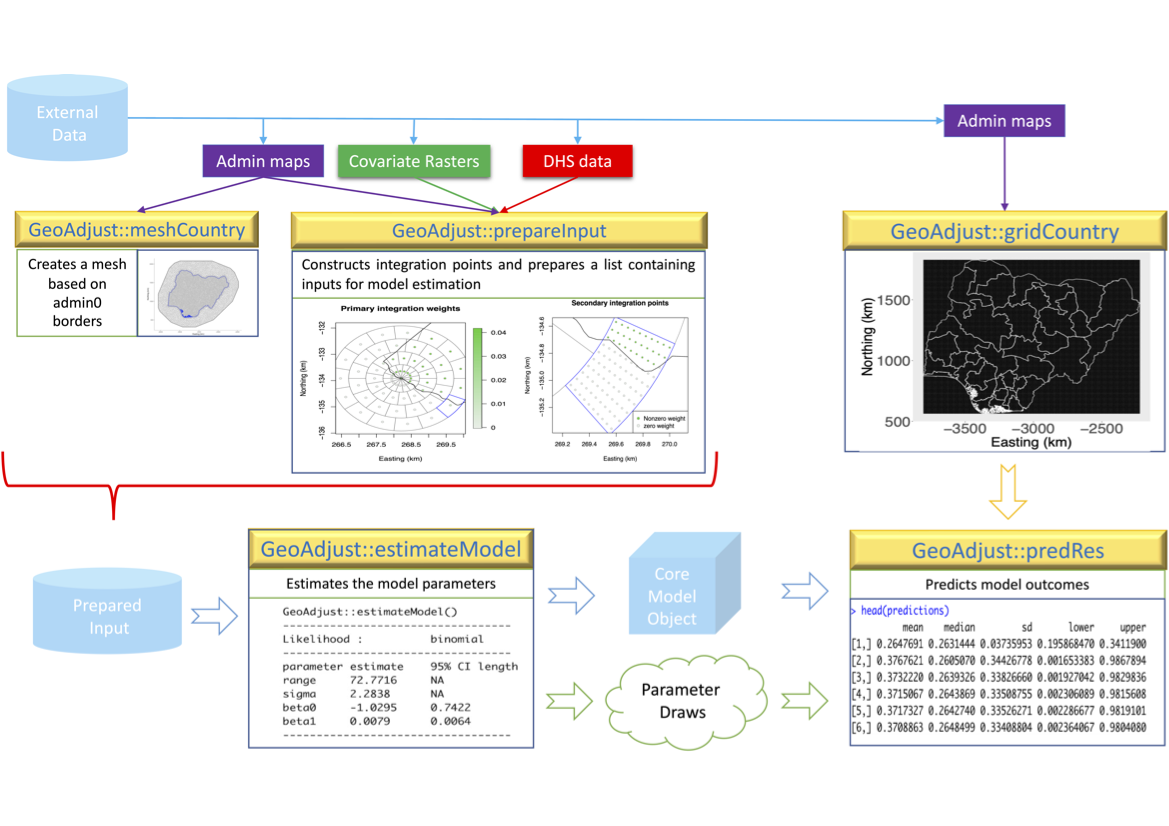}
\caption{\href{https://github.com/umut-altay/GeoAdjust-package}{\tt GeoAdjust} R-package workflow}
\label{fig:workflow}
\end{figure}

\href{https://github.com/umut-altay/GeoAdjust-package}{\tt GeoAdjust} hides the  complicated and technical steps in the algorithm from the user, to make the adjustment for jittering widely accessible.  Figure \ref{fig:workflow} illustrates the structure of \href{https://github.com/umut-altay/GeoAdjust-package}{\tt GeoAdjust}, and how various data inputs are processed through the package workflow. The main functionality of the package is described in subsections.

\subsection{Triangulation and mesh generation for the region under study}
In \href{https://github.com/umut-altay/GeoAdjust-package} {\tt GeoAdjust}, the GRF $u(\cdot)$ is approximated  using the so-called SPDE approach. This requires the construction of a constrained refined Delaunay triangulation (CRDT), in other words a mesh, over the country of interest. The approximated spatial field can then be projected from the mesh nodes to the cluster centers, by projector matrices \citep{Lindgren:etal:11}. The function \texttt{meshCountry}  creates a triangulation mesh based on the national borders.
It has two key arguments: \texttt{max.edge}  is a vector of two values, where its first and second elements represent the largest allowed triangle edge lengths for the inner and outer mesh, respectively, and \texttt{offset} stands for the extension distance outside the country borders. 

\subsection{Input data preparation}
The integration in Equation \eqref{eq:numInt} is done numerically and we need a set of integration points around each jittered survey cluster center. \href{https://github.com/umut-altay/GeoAdjust-package}{\tt GeoAdjust}  specifies the cluster center itself as the first integration point and builds either 5 or 10 rings around it, depending on whether it is located in an urban or a rural stratum, respectively. Each ring contains a set of 15 angularly equidistant primary integration points.

\begin{figure}
\centering
\includegraphics[width=.49\textwidth]{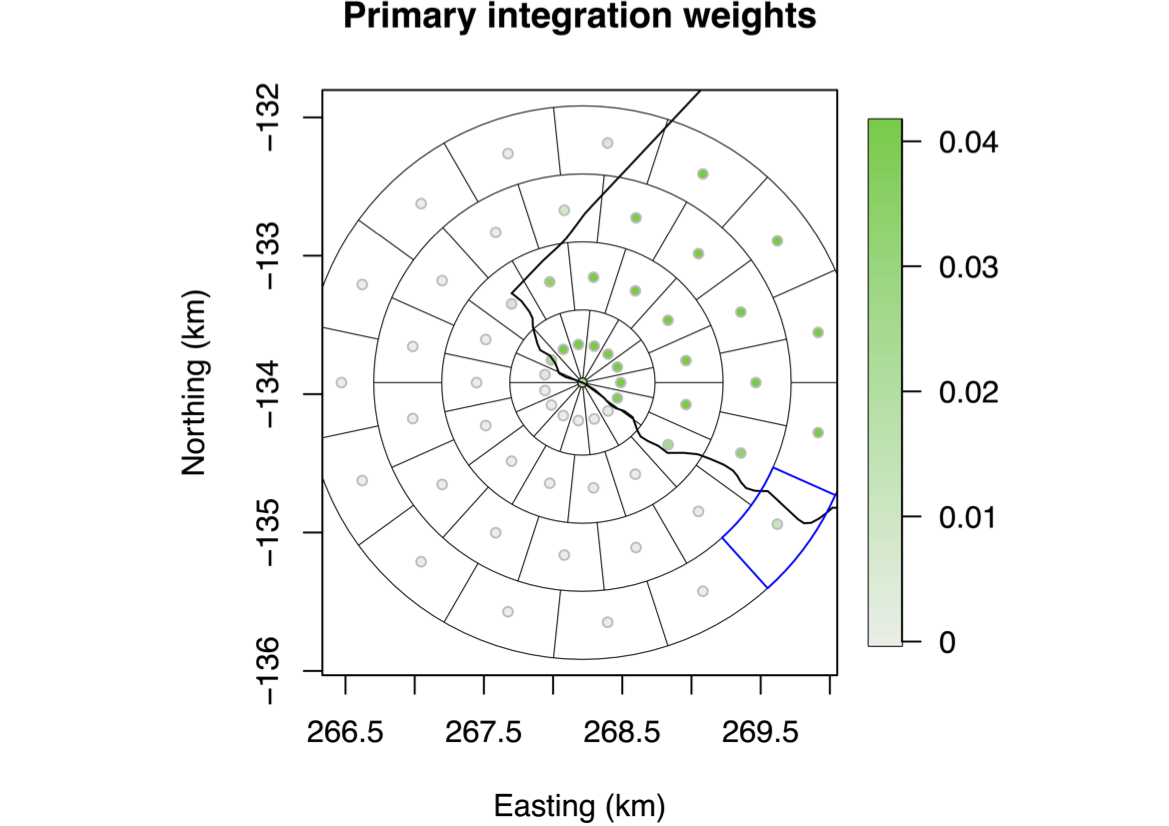} \includegraphics[width=.49\textwidth]{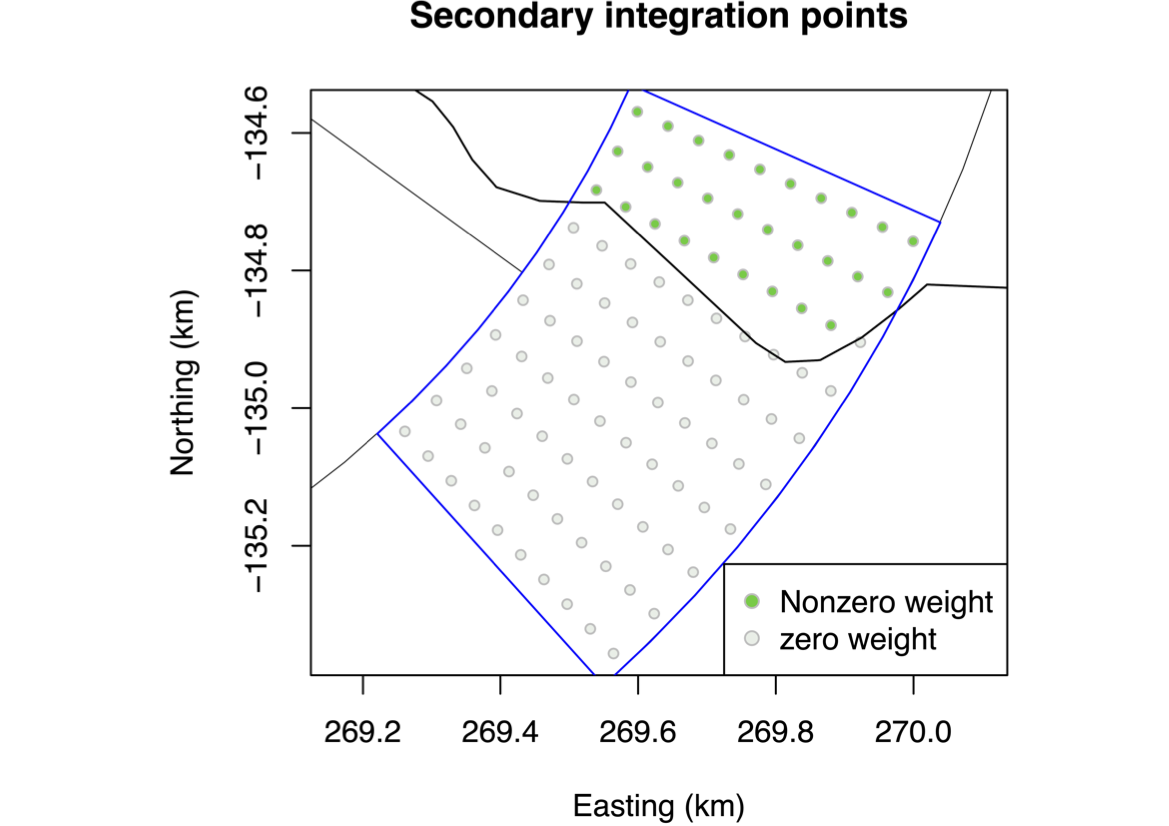}
\caption{Illustration of primary (left) and secondary (right) integration weights for one cluster from Kenya 2014 DHS household survey.}
\label{fig:intPtIllustration}
\end{figure}

The first 5 rings are called the "inner rings" and the points located within them are weighted equally. The additional 5 rings are built for the rural cluster centers and are called the "outer rings". The primary integration points within them are also assigned equal weights, which are smaller than the ones assigned for the points within the inner rings. If an observed cluster location is closer to the nearest subnational border than the maximum jittering distance, the method deploys a set of secondary integration points, each with an associate primary integration point, and assigns zero weight to any that are across the border. The associated primary integration point weights are adjusted accordingly. Figure \ref{fig:intPtIllustration} shows the primary and secondary integration points and the corresponding integration weights, for a single cluster from the Kenya 2014 DHS household survey with observed location close to an administrative boundary. The supplementary materials section of \citep{altay2022accounting} provides a detailed mathematical explanation about the procedure.

 The function \texttt{prepareInput} creates the set of integration points and the corresponding weights with respect to the urbanization strata, and constructs the urban and rural design matrices by extracting the covariate values at the coordinates of each integration point. 
The function returns a list containing the strata-wise design matrices and response vectors, together with the sparse matrix components of the SPDE model, and strata-wise projector matrices. Usage of the function will be shown based on the NDHS-2018 survey in Section \ref{sec:NigeriaExample}.

\subsection{Model parameter estimation}
The input list returned by \texttt{prepareInput} function consists of the elements that will be processed by the autodifferentiation feature of TMB as implemented in the \texttt{estimateModel} function. 
The function \texttt{estimateModel} saves the users from writing a complex C++ code to run TMB, and allows estimating model parameters by setting a small number of arguments. The function is flexible and allows different prior choices for the model components, via its argument called \texttt{priors}.

The function \texttt{estimateModel}  utilizes the C++ script of TMB, which integrates out the unknown true coordinates by computing the contribution of each integration point to the joint negative log-likelihood. 
Internally, once the TMB function \texttt{MakeADFun} constructs the core model object \citep{kaskr2022}, \texttt{estimateModel} uses the \texttt{optim} function to optimize it. Afterwards, \texttt{estimateModel} extracts the estimated model parameters from the optimized core model object, and draws samples of size, that is controlled by the argument \texttt{n.sims} of \texttt{estimateModel}function, for each covariate effect. The function draws samples of size \texttt{n.sims} for the spatial random effect coefficients for each mesh node as well. The samples for the intercept and the covariate effect sizes are then used for constructing the 95\% credible interval lengths as the measure of uncertainty corresponding to the estimated parameters.

The function \texttt{estimateModel} returns a list of four elements. The list contains a data frame of the estimated model parameters, together with the optimized core model object, a matrix containing the drawn samples of size \texttt{n.sims} for the covariate effect sizes and the random effect coefficients, and information about the type of the likelihood. The core model object and the drawn samples can then be passed to the function \texttt{predRes} to generate predictions at a set of prediction locations.

\subsection{Prediction grid construction}
Once the model parameters are estimated, the model can be used for predicting the model outcomes at a new set of locations. The function \texttt{gridCountry} in \href{https://github.com/umut-altay/GeoAdjust-package}{\tt GeoAdjust} helps with the construction of a set of prediction points. The function creates a raster of a preferred resolution within the bounding box of the national level shape file, extracts the coordinates of the cell centers and returns them together with the raster, as the elements of a list. A code example about the implementation of this function is given in Section \ref{sec:NigeriaExample}.

\subsection{Prediction}
Obtaining predictions at a new set of locations the function {\tt predRes} requires the optimized core model object, drawn samples of the parameters and the random effect coefficients, triangulation mesh, a list of covariate rasters, coordinates of the prediction locations and an argument called "flag" as input arguments. The argument "flag" is used for passing the likelihood type into the function. The integers 0, 1 and 2 indicate the Gaussian, binomial and Poisson likelihoods, respectively, and the function deploys the corresponding link function as outlined before. The package allows the use of any number of covariates, as long as they are in a geospatial raster layer format. The covariates can be passed into the function within a single list. The function will extract the values from each one of them at the prediction locations and form a design matrix.  

Internally, \texttt{predRes} combines the sampled covariate effect sizes and the random effect coefficients with the design matrix and forms one model per sample, \texttt{n.sims}
models in total. Each model predicts outcomes across the set of prediction locations. Finally, the function calculates the mean, median, standard deviation, and the upper and lower bounds of 95\% credible intervals of predictions for each prediction point. These results are returned in a matrix with a number of rows equal to the number of prediction points, and 5 columns. 

\section{DHS data acquisition}
The data sets of DHS household surveys are semi-public, but access to them requires application for permission. A step by step guidance to the application procedure can be found in \url{https://dhsprogram.com/data/new-user-registration.cfm}. The application requires a brief project description  explaining why the data set is needed and how it will be used within the provided project framework. Sharing the data sets with each one of the collaborating researchers requires permission as well. Once the permission is granted, the applicant receives a letter via email, which clearly states the content of the permission.

\section{Preprocessing DHS data}
Geospatial analysis of DHS household surveys usually require processing the individual level responses together with the cluster level information. Prior to using \href{https://github.com/umut-altay/GeoAdjust-package}{\tt GeoAdjust}, the DHS data needs to be preprocessed and certain variables need to be extracted. The package uses the clusterID, cluster center coordinates (both in degrees and in kilometers), administrative area names in which the clusters are located in, and the Gaussian, binomial or Poisson outcome variable aggregated from the individual responses into each cluster center. 
Two administrative border shape files are used in the analysis. One of them is the shape file that contains the national (admin0) level borders of the country of interest. The second one contains the subnational administrative level borders which are respected while  jittering. Shape files of various administrative levels for different countries can be obtained from the  website of "the Database of Global Administrative Areas (GADM)" (\url{https://gadm.org/data.html}). Once downloaded, the files can be read into R as "SpatialPolygonsDataFrame" objects, see Listing~\ref{lst:generalExample} for an example. 
Finally, if any raster- and distance-based covariates will be included in the model, they need to be read into R as separate raster layers. The R code for reading and further processing the administrative borders shape files and the covariate rasters will be shown in Section \ref{sec:NigeriaExample}. 
Once the preprocessing is done, in other words, all the external data files are read in and the variables of interest are extracted and stored in a data frame, the functions \texttt{meshCountry} and \texttt{prepareInput} can be used (see Section~\ref{sec:NigeriaExample}).

\subsection{Reading data \label{sec:readingData}}
The individual responses and the cluster information are often contained in separate files in different formats. The survey responses are collected via questionnaires and the answers of the participants to each question are stored under the corresponding variable names within one large data file. Descriptions of the variables can be found from DHS recode manuals such as \citet{fund2018demographic} and the response of interest can be aggregated into the cluster centers that is stored in the cluster level data file. The aggregation step must be adapted to the application. 
Listing~\ref{lst:generalExample} shows how to read the DHS data into R and set it in the working environment. 

\begin{lstlisting}[language=R, caption=Loading DHS data into R., label={lst:generalExample}]
library(haven)
library(rgdal)
# Reading DHS data :
# path1 : the full path to the individual level data file (.DTA)
# path2 : the path to the folder where the cluster level file (.shp) is located 

# individual level data (individual survey responses) :
individualData = read_dta(path1) 

# cluster level data (clusterID, cluster center coordinates, strata, etc.)
corList = readOGR(dsn = path2,layer = "file name")

# extract cluster level information:
smallGeo = data.frame(clusterIdx = corList$DHSCLUST, 
                      urban = corList$URBAN_RURA,
                      long = as.vector(corList@coords[,1]), 
                      lat = as.vector(corList@coords[,2]),
                      admin1 = corList$ADM1NAME)

#  extract individual level information:
myData = data.frame(clusterIdx = individualData$v001,   
                    variable1  = individualData$v1,           
                    variable2  = individualData$v2)  
\end{lstlisting}

\subsection{Example for Nigeria \label{sec:NigeriaExample}}

\begin{figure}
\centering
\includegraphics[width=.43\textwidth]{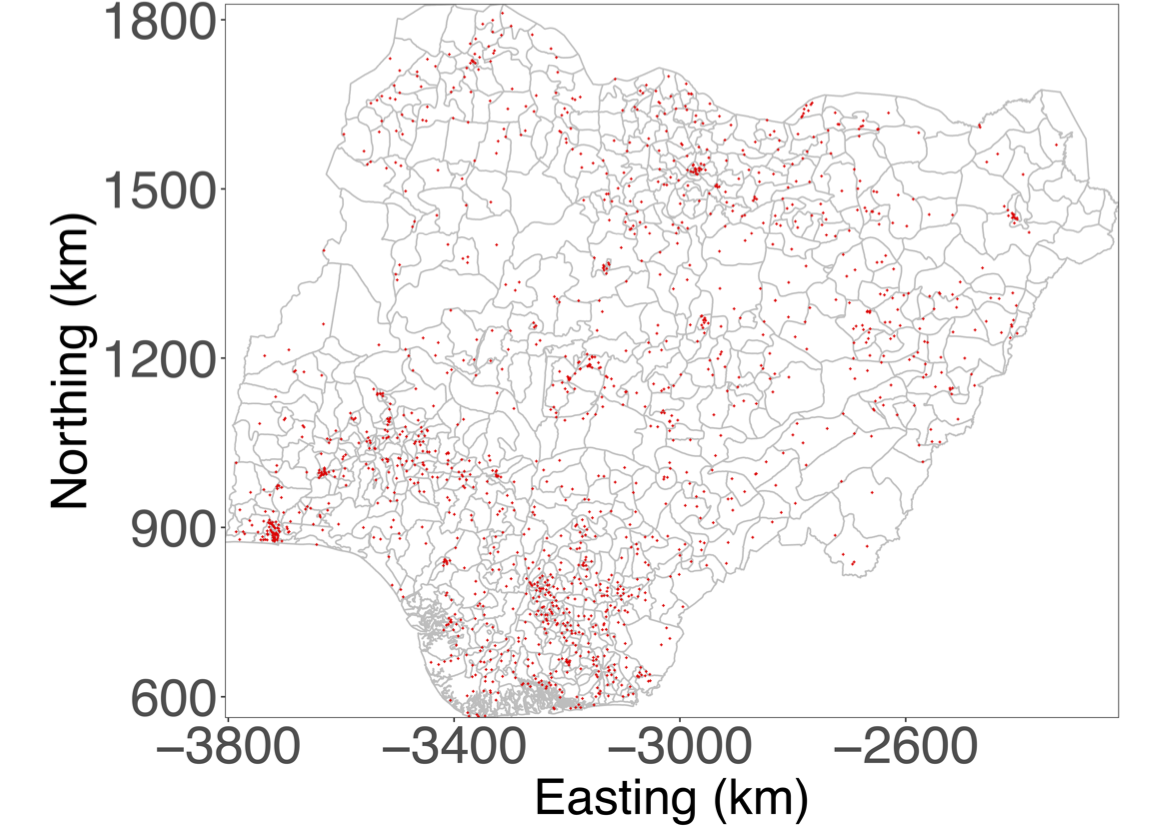} \includegraphics[width=.43\textwidth]{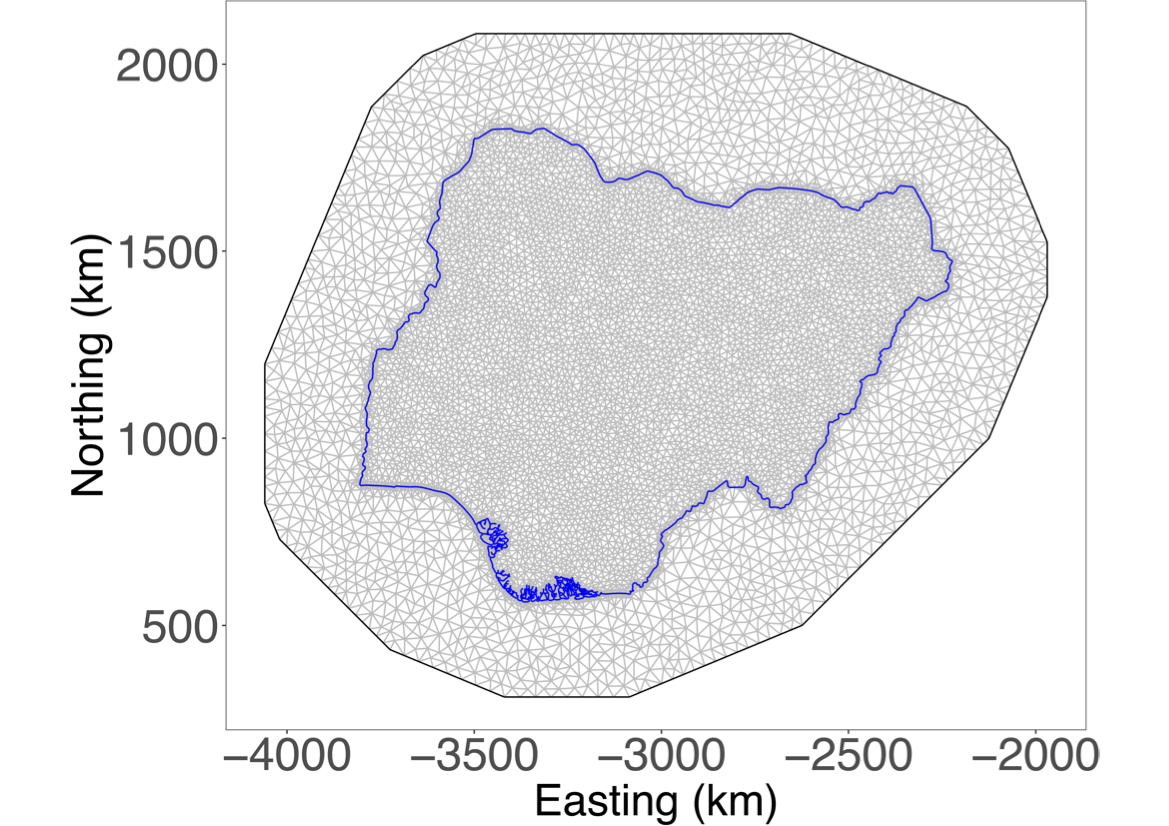}
\caption{Nigeria subnational level map (left) and the triangulation mesh (right). The red points represent the jittered cluster centers.}
\label{fig:meshAndCountry}
\end{figure}

 This section shows extracting and merging individual and cluster level data, based on Nigeria DHS 2018 (NDHS-2018) household survey. The example code in this section considers as outcome the completions of secondary education among 20-39 years old women in Nigeria. Population density is used as the only covariate and the corresponding raster file (Nga\_ppp\_v2c\_2015.tif) can be downloaded from WorldPop \citep{pop}. The geography of Nigeria and the locations of the clusters are shown in the left-hand side panel of Figure \ref{fig:meshAndCountry}.
 
 This example will use the model 
 \begin{align}
\label{eqn:NigeriaModel}
\begin{split}
y_c | r_c,n_c &\sim \text{Binomial}(n_c, r_c), \quad \boldsymbol{s}_c|\boldsymbol{s}_c^*\sim \pi_{\mathrm{Urb}[c]}(\boldsymbol{s}_c|\boldsymbol{s}_c^*),\\
    r_c  &= r(\boldsymbol{s}_c^*) = \mathrm{logit}^{-1}( \eta(\boldsymbol{s}_c^*)),
\end{split}
\end{align}
where $y_c$ is the number of women who completed secondary education, $n_c$ is the number of women interviewed, and $r_c$ denotes the risk in cluster $c$, for $c = 1, \ldots, C$. The spatially varying risk $r(\cdot) = \mathrm{logit}^{-1}(\eta(\cdot))$ is  modelled through the linear predictor 
\[
\eta(\boldsymbol{s}^*) = \beta_0 +x(\boldsymbol{s}^*)\beta_1 + u(\boldsymbol{s}^*), \quad \boldsymbol{s}^*\in\mathcal{D},
\]
where $\beta_0$ is the intercept, $x(\boldsymbol{s}^*)$ is the population density, $\beta_1$ is the effect of population density, and $u(\cdot)$ is the Matérn GRF with smoothness $\nu = 1$.
 
 The shape file for Nigeria includes a large lake on its north-eastern corner. Lakes do not have any DHS household clusters within them, therefore it does not make sense to make any predictions at locations that are within the lake. Accordingly, we remove the polygon that corresponds to the lake from the admin2 level shape file. Listing~\ref{lst:NigeriaExample1} shows reading the administrative area shape files and DHS data and extracting the variables of interest based on the file and variable names of NDHS-2018. 
 
\begin{lstlisting}[language=R, caption={Data preprocessing: loading administrative shapefiles, DHS data, and covariate data into R for the Nigeria example.}, label={lst:NigeriaExample1}]
# reading admin0 and admin2 shape files :
admin0 = readOGR(dsn = "dataFiles/gadm40_NGA_shp",
                            layer = "gadm40_NGA_0")

admin2 = readOGR(dsn = "dataFiles/gadm40_NGA_shp",
                     layer = "gadm40_NGA_2")
                     
# remove the lake
admin2 = admin2[-160,] # Nigeria map has a large lake 
                       # The lake corresponds to polygon 160
                       
# reading DHS data :
corList = readOGR(dsn = "dataFiles/DHS/NG_2018_DHS_02242022_98_147470/NGGE7BFL",
                              layer = "NGGE7BFL")
educationData = read_dta("NGIR7BDT/NGIR7BFL.DTA")

# extract cluster level information:
smallGeo = data.frame(clusterIdx = corList$DHSCLUST, 
                      urban = corList$URBAN_RURA,
                      long = as.vector(corList@coords[,1]), 
                      lat = as.vector(corList@coords[,2]),
                      admin1 = corList$ADM1NAME)

#  extract individual level information:
myData = data.frame(clusterIdx = educationData$v001,  # cluster ID
                    age = educationData$v012,                # age 
                    secondaryEducation = educationData$v106) # v106 
                    #v106 : highest education level 
                    # 0  : no education
                    # 1  : primary
                    # 2  : secondary
                    # >2 : higher

# reading the covariate raster:
library(raster)
r = raster::raster("Nga_ppp_v2c_2015.tif")
\end{lstlisting}

Once the external data files are read into R, the data needs to be organized with respect to the content of the research. Accordingly, the individual survey answers contained in the data frame "\texttt{myData}" are first subsetted with respect to the age interval that we are interested in (20-39), and then merged with the cluster level information in the data frame "\texttt{smallGeo}". The merged data are then aggregated into the cluster centers. These steps can be followed through Listing~\ref{lst:NigeriaExample2}.

\begin{lstlisting}[language=R, caption=Data preprocessing: subsetting and aggregating., label={lst:NigeriaExample2}]
# subset data with respect to the age interval of interest:
myData = subset(myData, age <= 39 & age >=20)

# number of 20-39 years old women who completed secondary education in each household
myData$ys = (myData$secondaryEducation>=2)+0

# merge the cluster level data with the subsetted individual level data,
# with respect to the cluster ID:
myData = merge(myData, smallGeo, by = "clusterIdx")

# add number of trials (for binomial response)
myData$Ntrials = 1

# aggregate the survey responses to the cluster centers
answers_x = aggregate(myData$ys,
                      by = list(clusterID = myData[, 1]),
                      FUN = sum)

answers_n= aggregate(myData$ys,
                     by = list(clusterID = myData[, 1]),
                     FUN = length)

# merge
answers_joint = merge(answers_x, answers_n, by="clusterID")

# now we have the total number of women participants within the relevant age interval (ns), 
# for each cluster. We also have the number of women among those who completed their secondary education (ys)
colnames(answers_joint) = c("clusterID", "ys", "ns")
\end{lstlisting}

The main variables that are needed for the analysis are the ID numbers and coordinates of the cluster centers (both in degrees and in kilometers), their urbanicity stratum, and the aggregated response variable values. Accordingly, these are collected into a main data frame as in Listing~\ref{lst:NigeriaExample3}.

\begin{lstlisting}[language=R, caption=Data preprocessing: collecting data into a data frame., label={lst:NigeriaExample3}]

# initial data frame
nigeria.data = data.frame(clusterID = corList@data[["DHSCLUST"]], long = as.vector(corList@coords[,1]), lat = as.vector(corList@coords[,2]))

# add ys and ns
nigeria.data = merge(nigeria.data, answers_joint, by="clusterID", all=T)

# add strata:
nigeria.data$urbanRuralDHS = corList@data[["URBAN_RURA"]]

# add coordinates in kilometers
nigeria.data$east = rep(NA, length(nigeria.data$long))
nigeria.data$north = rep(NA, length(nigeria.data$long)) 

nigeria.data[,c("east", "north")] = convertDegToKM(nigeria.data[,c("long", "lat")])
\end{lstlisting}

The DHS jittering scheme is implemented by respecting various levels of administrative borders in different countries. The function \texttt{prepareInput} creates the integration points and considers their proximity to the respected level of administrative borders to decide if a secondary set of points should also be deployed. In NDHS-2018, jittering is done by respecting the second administrative level borders in Nigeria. It is important to be sure that the admin2 level areas that each cluster center is located within can be identified, in other words, each cluster center matches with one of the areas. This is the information that will lead the function \texttt{prepareInput} to evaluate the proximity of each individual integration point to the borders of the corresponding particular administrative area. Accordingly, the cluster centers that do not match with any admin2 areas need to be dropped as shown in Listing~\ref{lst:NigeriaExample4}.

\begin{lstlisting}[language=R, caption=Data preprocessing: final adjustments., label={lst:NigeriaExample4}]
# jittering is done by respecting admin2 borders in Nigeria. 
# see if there are  cluster centers that doesn't match with any of the admin2 areas:

# first, add polygon IDs (some shape files may have it already) :
admin2@data[["OBJECTID"]] =1:774 # normally 775, we removed one (the lake)
                                 # this number might be different in other   countries

# the cluster coordinates:
latLon = cbind(nigeria.data[,"long"], nigeria.data[,"lat"])
colnames(latLon) = c("long", "lat")

# make a SpatialPoints object
latLon = SpatialPoints(latLon, proj4string=CRS("+proj=longlat +datum=WGS84 +no_defs"), bbox = NULL)

# see if the points (cluster centers) are within the polygons (admin2 areas) :
check1 <- over(latLon, admin2, returnList = FALSE)

# drop the rows which don't match with none of the admin2 areas. 
# we will need them to match while creating the integration points later on.

# see which ones don't return a match :
# which(is.na(check1$NAME_2))  # see the rows that don't match:
# [1]   48  122  205  848  857 1116 1122 1287 1328

# drop the corresponding rows from the main data set :
nigeria.data = nigeria.data[-c( 48, 122, 205, 848,  857, 1116, 1122, 1287, 1328),]
\end{lstlisting}

Besides creating the integration points, the \texttt{prepareInput} function constructs the SPDE components and includes them in its returning list. The function implements this based on the triangulation mesh. This last step before running \texttt{prepareInput} is illustrated in Listing~\ref{lst:NigeriaExample5}.

\begin{lstlisting}[language=R, caption=Data preparation: constructing a triangulation mesh with the \texttt{meshCountry()} function., label={lst:NigeriaExample5}]
# transform admin0 into kilometers 
proj = "+units=km +proj=utm +zone=37 +ellps=clrk80                                +towgs84=-160,-6,-302,0,0,0,0 +no_defs"

admin0_trnsfrmd = spTransform(admin0,proj)

library(GeoAdjust)
# construct the mesh
mesh.s = meshCountry(admin0 = admin0_trnsfrmd, 
                     max.edge = c(25, 50), offset=-.08)
\end{lstlisting}

Figure \ref{fig:meshAndCountry} shows the subnational (admin2 level) borders within Nigeria, together with the triangulation mesh that is constructed based on the national borders. 

The function \texttt{prepareInput} saves the package user from various detailed long coding tasks and reduces the whole process into setting just couple of arguments. The covariates in the model need to be passed into the function as raster layers within a list, via the argument "\texttt{covariateData}". The function needs the response variable values as well, so that it can construct the response vectors for the corresponding urban and rural integration points, separately. This example on NDHS-2018 has a binomial response. Accordingly, we need to pass both the aggregated binomial trials (ns) and the corresponding aggregated binomial successes (ys) for each cluster center. Here, ns represents the number of 20-39 years old women per cluster, and ys is the number of women who reported that their secondary education is completed, amongst them. If the response was Gaussian or Poisson, then the list passing to the "\texttt{response}" argument would only contain ys either as the Gaussian responses or the Poisson counts, respectively. Similarly, the likelihood type should be set via the argument "\texttt{likelihood}", so that the function processes the other arguments accordingly. Here, the values 0, 1 and 2 indicates that either Gaussian, binomial or Poisson likelihood is used in the model, respectively. The argument "\texttt{jScale}" sets the scaling of the default DHS maximum jittering distances. The function \texttt{prepareInput} multiplies the default distances by the value set to the argument, and evaluates the approximity of the primary integration points to their corresponding administrative area borders based on the scaled distances. The value 1 indicates that the default distances are in use. Different values can be set to this argument in order to experiment with them. Listing~\ref{lst:NigeriaExample6} shows how the arguments can be set and how the function can be used.

\begin{lstlisting}[language=R, caption=Data preparation: preparing a list of input objects with the \texttt{prepareInput()} function., label={lst:NigeriaExample6}]
# read the covariate raster
library(raster)
r = raster::raster("Nga_ppp_v2c_2015.tif")

# the response variable
response = list(ys = nigeria.data$ys, # number of binomial successes
               ns = nigeria.data$ns)  # number of binomial trials
               
# cluster center coordinates in kilometers
locObs =cbind(nigeria.data[["east"]], nigeria.data[["north"]])

likelihood = 1 # binomial likelihood
               # (set 0, 1 or 2 for Gaussian, binomial or Poisson)
jScale = 1     # the maximum DHS jittering distances 
               # can be scaled using this argument
               # 1 corresponds to the default DHS jittering


adminMap = admin2  # jittering is done by respecting admin2 
                    # borders in Nigeria. This may be different
                    # for other countries. In Kenya, admin1 borders 
                    # are respected instead.

inputData = prepareInput(response=response, locObs=locObs, 
                         likelihood = likelihood, 
                         jScale=jScale, 
                         urban = nigeria.data$urbanRuralDHS, 
                         mesh.s = mesh.s, adminMap=adminMap, nSubAPerPoint=10, nSubRPerPoint = 10,
                         covariateData=list(r=r))
\end{lstlisting}

The content of the input list created by function \texttt{prepareInput} is shown in Listing~\ref{lst:NigeriaExample7}. The list contains the response vectors (both ns and ys in binomial case), design matrices and the projector matrices all created separately for the urban and rural strata. The function \texttt{prepareInput} breaks the vectors ns and ys into the urban and rural vectors \texttt{num\_iUrban}, \texttt{num\_iRural}, \texttt{y\_iUrban} and \texttt{y\_iRural} with respect to the urban and rural integration points. The other elements are the coordinates of the urban and rural integration points, corresponding urban and rural integration weights, SPDE components and the likelihood and normalization flags.

\begin{lstlisting}[language=R, caption=Data preparation: the input list., label={lst:NigeriaExample7}]
The final input data list contains the following elements:
 inputData <- list(num_iUrban,  # Total numb. of urban obs.
               num_iRural,  # Total numb. of rural obs.
               num_s, # num. of vertices in SPDE mesh
               y_iUrban, # urban obs in the cluster
               y_iRural, # rural obs in the cluster
               n_iUrban,  # urban exposures in the cluster
               n_iRural,  # rural exposures in the cluster
               n_integrationPointsUrban,#num.of urb.int.pts.
               n_integrationPointsRural,#num.of rur.int.pts.
               wUrban = wUrban, # urban weights
               wRural = wRural, # rural weights
               X_betaUrban = desMatrixJittUrban, # urb. des. mat.
               X_betaRural = desMatrixJittRural, # rur. des. mat.
               M0, #=spde[['param.inla']][['M0']], 
               M1, #=spde[['param.inla']][['M1']], 
               M2, #=spde[['param.inla']][['M2']], 
               AprojUrban,  # Projection matrix (urban)
               AprojRural,  # Projection matrix (rural)
               options = c(1, ## if 1, use normalization
                           1), ## if 1, run adreport
               flag1 = 1, # normalization flag.
               flag2 = flag2, #(0/1/2 for Gaussian/Binomial/Poisson)
  )
  )
\end{lstlisting}

\section{Model estimation and gridded spatial prediction} 

\subsection{Estimation}
The function \texttt{estimateModel} utilizes the \texttt{MakeADFun} function of TMB to construct a list we will refer to as the core model object, containing the objective functions with derivatives \citep{tmb}, \citep{kaskr2022}. Then \texttt{estimateModel} uses \texttt{optim} to optimize the core model object and estimate the model parameters, without requiring the user to write any C++ code. The main argument of the function is a list called ``\texttt{data}'', referring to the input list that has just been created above. The argument \texttt{nNodes} refers to the number of nodes that the triangulation mesh has. The remaining two other arguments are called \texttt{options} and \texttt{priors}.

The argument \texttt{options} specifies in which of the two model components, namely, the random effect and covariates, jittering should be accounted for. Jittering adjustment can be turned on and off either in the random effect or in covariates or both,  by setting the values of ``\texttt{random}'' and ``\texttt{covariates}'' to 1 or 0, respectively.

The argument \texttt{priors} allows the user to specify the parameters of the Gaussian prior for covariate effect sizes, and of the penalized complexity (PC) priors for the spatial range and marginal variance. These values can be passed into the function as a list of six elements, namely, ``\texttt{beta}'', ``\texttt{range}'', ``\texttt{Uspatial}'', ``\texttt{alphaSpatial}'', ``\texttt{UNugget}'', and ``\texttt{alphaNug}''. 
The element beta needs to be a vector of length two. The first and the second elements of the vector beta are the mean and the standard deviation of the Gaussian priors that are assigned for the intercept and the covariate effect sizes. ``\texttt{range}'' is the \emph{a priori} median range, and ``\texttt{USpatial}'' is the upper ``\texttt{alphaSpatial}'' percentile of the marginal standard deviation, and ``\texttt{UNugget}'' and ``\texttt{alphaNug}'' are the hyperparameters for the PC-prior on the nugget variance. The hyperparameters ``\texttt{UNugget}'' and ``\texttt{alphaNug}'' pass into the function as 1 and 0.05, by default, but they are only used in the calculations when the likelihood is Gaussian. The package user is free to fix them to other values as well. Listing~\ref{lst:NigeriaExample8} shows how the \texttt{estimateModel} function can be used. The argument ``\texttt{n.sims}'' controls the number of samples that will be drawn for each model parameter and each random effect coefficient.

\begin{lstlisting}[language=R, caption=Estimating model parameters: using the \texttt{estimateModel()} function., label={lst:NigeriaExample8}]
# number of nodes in the mesh:
nNodes = mesh.s[['n']]

# estimating the parameters
est = estimateModel(data = inputData, 
      nNodes = nNodes,
      options = list(random = 1, covariates = 1), # account for jittering in random and covariate effects
      priors = list(beta = c(0,1), 
        range = 114, 
        USpatial = 1, alphaSpatial = 0.05, UNugget = 1, alphaNug = 0.05), n.sims = 1000)
\end{lstlisting}

\texttt{estimateModel} returns a list of four elements (see Listing~\ref{lst:NigeriaExample9}). Two of them, namely, \texttt{obj} and \texttt{draws} will be passed into the \texttt{predRes} function for predictions on a new prediction grid. The element \texttt{obj} is the optimized core model object. The element \texttt{draws} contains \texttt{n.sims} draws for the effects of covariates and the random effect. The element \texttt{likelihood} indicates the likelihood type that is used in the model construction. Finally, the last element \texttt{res} contains the estimated model parameters and the lengths of 95\% credible intervals, which are constructed using the sampled values in \texttt{draws}. The credible interval lengths are calculated within the \texttt{estimateModel} function as the difference between the 97.5\%  and 2.5\%  quantiles of the drawn samples for the corresponding parameter estimate. The result object \texttt{res} does not contain \texttt{CI\_Length} values for the range and the marginal variance, as the inference is empirical Bayesian where these parameters are estimated to fixed values. The model estimates can be printed in a tidy way as in Listing~\ref{lst:NigeriaExample9}
\begin{lstlisting}[language=R, caption=Estimating model parameters: output., label={lst:NigeriaExample9}]
# the output of estimateModel() function:
names(est)
[1] "res"        "obj"        "draws"      "likelihood"

print(est)

GeoAdjust::estimateModel() 
----------------------------------
Likelihood :          binomial
----------------------------------
parameter estimate    95% CI length
range     72.7716     NA
sigma     2.2838      NA
beta0     -1.0295     0.7488
beta1     0.0079      0.0064
----------------------------------
\end{lstlisting}

\subsection{Prediction grid}
\href{https://github.com/umut-altay/GeoAdjust-package}{\tt GeoAdjust} provides the \texttt{gridCountry} function to create a grid of prediction points with respect to the national borders of the country of interest. The function has  two arguments. The first argument \texttt{admin0} is the SpatialPolygonsDataFrame object containing the national borders. The second argument \texttt{res} indicates the resolution in kilometers. Internally, the function first creates a raster within the bounding box of the \texttt{admin0}  object and with respect to the chosen resolution. Afterwards, it extracts the coordinates of the cell centroids and constructs a data frame containing the cell centroid coordinates both in kilometers and degrees. Finally, the function returns the coordinates and the prediction raster within a list. Having the prediction raster is necessary to use the function \texttt{plotPred}, which internally utilizes \texttt{geom\_raster} from \texttt{ggplot2}, which is useful for plotting the predictions and the uncertainty across the country. Listing~\ref{lst:NigeriaExample10} shows how \texttt{gridCountry} function can be used.

\begin{lstlisting}[language=R, caption=Prediction: the \texttt{gridCountry()} function., label={lst:NigeriaExample10}]
# raster and the prediction coordinates:
predComponents = gridCountry(admin0 = admin0, res = 5)

names(predComponents)
[1] "loc.pred" "predRast"

# the prediction locations
loc.pred = predComponents[["loc.pred"]]

head(loc.pred)
       east    north     long      lat
1 -3803.287 1825.665 1.838939 13.27084
2 -3798.287 1825.665 1.875670 13.27712
3 -3793.287 1825.665 1.912420 13.28340
4 -3788.287 1825.665 1.949189 13.28967
5 -3783.287 1825.665 1.985977 13.29595
6 -3778.287 1825.665 2.022784 13.30222

> dim(loc.pred)
[1] 80201     4

predRast = predComponents[["predRast"]]
print(predRast)
class      : RasterLayer 
dimensions : 253, 317, 80201  (nrow, ncol, ncell)
resolution : 5, 5  (x, y)
extent     : -3805.787, -2220.787, 563.1654, 1828.165  (xmin, xmax, ymin, ymax)
crs        : +proj=utm +zone=37 +ellps=clrk80 +units=km +no_defs 

\end{lstlisting}

\subsection{Prediction}
The \texttt{predRes} function uses the core model object that is created within \texttt{estimateModel} to predict the model outcomes at a set of prediction locations. 
The function \texttt{predRes} uses two elements from the output list of \texttt{estimateModel}, namely the core model object (\texttt{est[["obj"]]}) and the matrix containing the sampled covariate effect sizes together with the sampled random effect coefficients for each mesh node (\texttt{est[["draws"]]}). 

In this example we use the cell center coordinates of the prediction raster which is just constructed by \texttt{gridCountry} function, but it is also possible for the package users to predict on a custom made grid or any other set of locations. Please note that, if the package user prefers to use a different location set, their coordinates need to be passed in kilometers as a matrix with the corresponding column names "east" and "north", respectively.

The function \texttt{prepareInput} used an argument  called "\texttt{covariateData}". It was a list containing the raster layers of each covariate. The purpose of the argument there was to extract the covariate values at the integration points and to create urban and rural design matrices. Similarly, \texttt{predRes} function uses the same argument with the same name and content, but here the function creates a design matrix by extracting the covariate values at the prediction locations. 

The argument \texttt{flag} takes one of 0, 1 or 2. The value of this argument indicates the type of the likelihood that the model includes. The values 0, 1 and 2 indicates the Gaussian, binomial and Poisson likelihoods, respectively. The function uses the value to decide which link function should be used. Listing~\ref{lst:NigeriaExample11} shows the usage of the function \texttt{prepareInput}.

\begin{lstlisting}[language=R, caption=Prediction: the \texttt{predRes()} function., label={lst:NigeriaExample11}]
predictions = predRes(obj = est[["obj"]] , predCoords = loc.pred,
                        draws = est [["draws"]] , nCov = nCov,
                        covariateData = covariateData,
                        mesh.s = mesh.s, flag = 1)

head(predictions)
          mean    median         sd       lower     upper
[1,] 0.2646259 0.2627431 0.03690252 0.196264273 0.3405225
[2,] 0.3742580 0.2584815 0.34268884 0.001753274 0.9855895
[3,] 0.3707435 0.2601383 0.33634250 0.002157952 0.9826507
[4,] 0.3687931 0.2615159 0.33290925 0.002405636 0.9801035
[5,] 0.3686910 0.2641275 0.33299652 0.002327227 0.9800920
[6,] 0.3677783 0.2623805 0.33203511 0.002436265 0.9803111

dim(predictions)
[1] 80201     5
\end{lstlisting}

\subsection{Plotting the predictions and uncertainty}

\begin{figure}
\centering
\includegraphics[width=.49\textwidth]{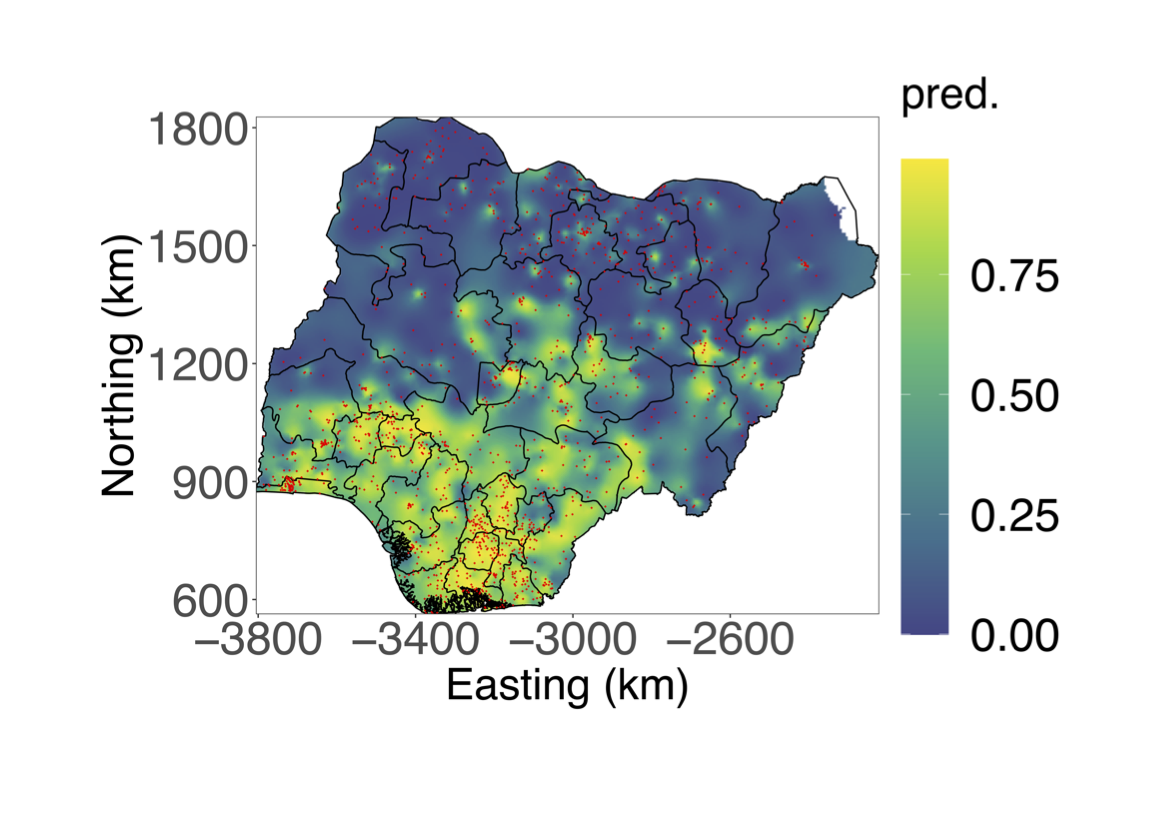} \includegraphics[width=.49\textwidth]{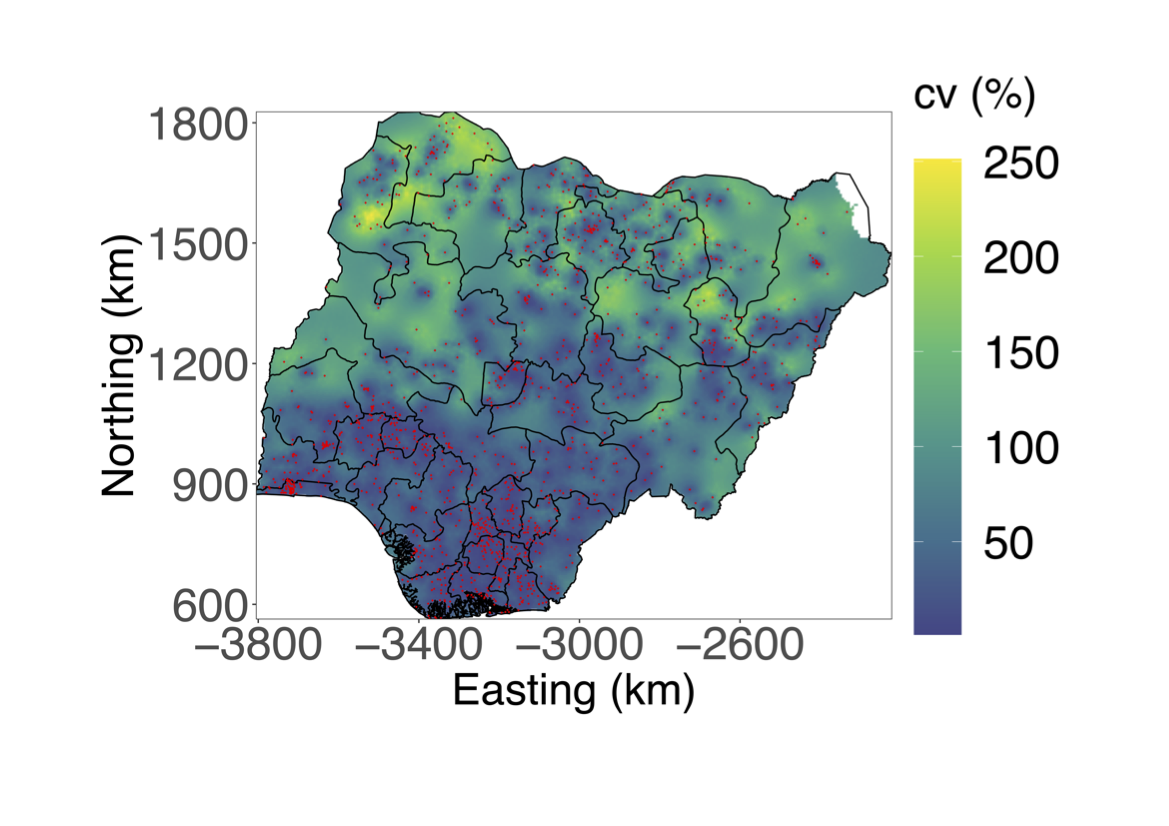}
\caption{Predicted risk (left) and the CVs (right). The red points indicate the example survey cluster centers.}
\label{fig:predPlot}
\end{figure}

This section shows how to plot the predicted values and the uncertainty accross the country map. We will use the predicted median values obtained from \texttt{predRes} function, as the point predictions. The plotted uncertainties will be the corresponding coefficient of variations calculated by $\frac{\sigma}{\mu} \times 100$, also obtained from \texttt{predRes} function. \href{https://github.com/umut-altay/GeoAdjust-package}{\tt GeoAdjust} uses the function \texttt{plotPred} to plot the predictions and the corresponding uncertainty accross the studied country. The first argument \texttt{pred} is the output of the function  \texttt{predRes} which is obtained in Listing~\ref{lst:NigeriaExample11}. The argument \texttt{predRaster} is the prediction raster that was constructed by \texttt{gridCountry} function in Listing~\ref{lst:NigeriaExample10}. The arguments \texttt{admin0}, \texttt{admin1} and \texttt{admin2} stand for the \texttt{SpatialPolygonsDataFrame} objects representing the national, first level subnational and second level subnational administrative borders of the corresponding country. There might be a need to leave some of the admin2 level polygons uncolored as we did here for the polygon 160 (the lake). Then the number of the polygon that needs to be excluded can be pass into the function through the argument \texttt{rmPoly}. The argument doesn't remove the polygon from the map. The function still plots the polygon within the map, but it doesn't assign colors anywhere within that polygon. The arguments \texttt{rmPoly} and \texttt{admin2} can be set to \texttt{NULL} if there is no such need. The administrative borders that are overlaid on the map by this function are the admin1 level borders (see Figure \ref{fig:predPlot}). The argument \texttt{locObs} indicates the observed locations, or in other words, the DHS cluster centers. The function plots these as red dots on to the map. The function returns a list containing two \texttt{ggplot} objects, representing the plots for the predictions and uncertainty accross the country of interest. Listing~\ref{lst:NigeriaExample12} shows how to use the function \texttt{plotPred}.

\begin{lstlisting}[language=R, caption=Plotting: preparation., label={lst:NigeriaExample12}]
admin1 = readOGR(dsn = "dataFiles/gadm40_NGA_shp",
                 layer = "gadm40_NGA_1")
                 
plotPred(pred = predictions, predRaster = predRast, admin0 = admin0,
          admin1 = admin1, admin2 = admin2, rmPoly = 160, locObs = locObs)
\end{lstlisting}

Figure \ref{fig:predPlot} shows the predicted risk and the corresponding uncertainty across Nigeria. Please note that since we assigned "NA" to the points that overlap with the lake, the north-east corner of the plots are not colored. This area is the area covered by the lake. This is specific to the geography of Nigeria and different features like this may need to be considered while plotting data on the maps of different countries.

\section{Summary}
\href{https://github.com/umut-altay/GeoAdjust-package}{\tt GeoAdjust} makes it easy to account for jittering, by isolating its user from complex code while still providing flexible control over the implementation. It is unique in a sense that it is the only package that specifically targets the positional uncertainty in the observed locations and also conveys a functionality emerging from a unique way of approaching to this problem. The package has a potential to be tested on and developed further for both areal and point referenced data from different areas involving the positional uncertainty and geomasking.

\bibliography{references}
\end{document}